\documentclass[a4paper]{article}

\usepackage{INTERSPEECH2022}

\usepackage{amsfonts} 
\usepackage{gensymb}  
\usepackage{multicol}
\usepackage{multirow}

\usepackage{amssymb}
\usepackage{pifont}

\usepackage{xpatch,xcolor}
\usepackage{hyperref}
\hypersetup{
    colorlinks=true,
    linkcolor=purple,
    filecolor=magenta,      
    urlcolor=purple,
}
\usepackage{bbm}

\usepackage{booktabs}

\ninept

\title{Representation Selective Self-distillation and wav2vec 2.0 Feature Exploration for Spoof-aware Speaker Verification}

\name{Jin Woo Lee$^1$, Eungbeom Kim$^{2}$, Junghyun Koo$^1$, and Kyogu Lee$^{1,2,3}$}
\address{
  $^{1}$ Department of Intelligence and Information, Seoul National University\\
  $^{2}$ Interdisciplinary Program in Artificial Intelligence, Seoul National University\\
  $^{3}$ AI Institute, Seoul National University
}

\email{\small\{jinwlee, eb.kim, dg22302, kglee\}@snu.ac.kr}

\begin{document}


\maketitle

\begin{abstract}
Text-to-speech and voice conversion studies are constantly improving to the extent where they can produce synthetic speech almost indistinguishable from bona fide human speech.
In this regard, the importance of countermeasures (CM) against synthetic voice attacks of the automatic speaker verification (ASV) systems emerges.
Nonetheless, most end-to-end spoofing detection networks are black-box systems, and the answer to what is an effective representation for finding artifacts remains veiled.
In this paper, we examine which feature space can effectively represent synthetic artifacts using wav2vec 2.0, and study which architecture can effectively utilize the space.
Our study allows us to analyze which attribute of speech signals is advantageous for the CM systems.
The proposed CM system achieved 0.31\% equal error rate (EER) on ASVspoof 2019 LA evaluation set for the spoof detection task.
We further propose a simple yet effective spoofing aware speaker verification (SASV) method, which takes advantage of the disentangled representations from our countermeasure system.
Evaluation performed with the SASV Challenge 2022 database show 1.08\% of SASV EER.
Quantitative analysis shows that using the explored feature space of wav2vec 2.0 advantages both spoofing CM and SASV.
\end{abstract}
\noindent\textbf{Index Terms}: speech anti-spoofing, spoofing aware speaker verification

\section{Introduction}
\label{sec:intro}

Automatic speaker verification (ASV) system aims to automatically verify the identity of individual speakers for given utterances.
To verify that the input sample is directly uttered by the target speaker, it is essential that ASV systems intrinsically reject any kind of spurious attempts, e.g., non-target utterances, synthetic speeches, or converted voices \cite{jung2022sasv}.
Although recent ASV systems have shown prominent results, such as achieving an equal error rates (EERs) less than 1\% in the \textit{in-the-wild} scenarios \cite{desplanques2020ecapa,thienpondt2021integrating}, there is room for improvement in being prepared for spoofing attacks \cite{todisco2019asvspoof}.

Text-to-speech (TTS) \cite{oord2016wavenet,wang2017tacotron,prenger2019waveglow} and voice conversion (VC) \cite{qian2019autovc,lee2021voicemixer,choi2021neural} studies have advanced to the extent that they could produce perceptually almost indistinguishable from bona fide human speech in terms of naturalness and similarity.
In order to deceive ASV systems, attackers aim to precisely mimic the target speaker's voice identity through replayed recordings, synthetic speeches, or cloned voices.
From this concern, various studies have been proposed to develop the ASV system to have robust reliability for spoofing \cite{todisco2019asvspoof}.

Early anti-spoofing studies were developed based on hand-crafted features such as spectral centroid frequency coefficients \cite{sahidullah2015comparison}, or constant-Q cepstral coefficients \cite{todisco2016new}.
However, it has empirically been shown that spoofing artifacts lie at the sub-band level \cite{sahidullah2015comparison,yang2019significance}, and that front-ends with high spectral resolution in the same band detect artifacts in reliably well \cite{jung2019replay,tak2020explainability}.
Since cepstral analysis compresses information on the entire spectral domain rather than emphasizing sub-band level features, the importance of research on advanced front-ends for spoof countermeasure networks compared to the conventional cepstral processing has emerged \cite{tak2020spoofing}.

Recent studies proposed end-to-end anti-spoofing systems that operate on raw waveforms \cite{jung2021aasist,tak2021endRAW,ma2021rw}.
These end-to-end models can also be viewed as a category of methods that replace the aforementioned feature extraction with neural nets with trainable parameters.
Several studies utilized sinc convolution-based network \cite{ravanelli2018speaker} as a front-end of anti-spoofing model \cite{jung2021aasist,tak2021endGAT}.

More recent studies investigated the effect of self-supervised front-ends as speech spoofing countermeasures \cite{tak2022automatic,wang2021investigating}.
Self-supervised learning has established itself as a powerful framework to learn general data representations from unlabeled data \cite{baevski2020wav2vec,devlin2018bert,liu2020mockingjay}.
Tak \textit{et al.} \cite{tak2022automatic} used wav2vec 2.0 model as front-end of existing AASIST \cite{jung2021aasist} countermeasure network.
They used output features of XLS-R \cite{babu2021xls} to improve the performance of AASIST.
Wang \textit{et al.} \cite{wang2021investigating} investigated several self-supervised models as feature extractors of the spoof detection task.
However, not many end-to-end anti-spoofing studies analyzed to find clues about the attributes that are important for finding artifacts.

Inspired by \cite{shah2021all}, where they investigated the most relevant features of each wav2vec 2.0 Transformer layer to the characteristics of speech signals, we examine features extracted from each layer of the wav2vec pre-trained model for the spoofed speech detection task.
Contrary to the approaches that
studied the use of self-supervised models as front-end of spoofing countermeasures, we seek to find answers to the following questions:
\begin{itemize}
    \item Which attribute is most related to the features advantageous for spoofing detection?
    \item How much simpler the back-end model can be behind the improved feature extractor?
    \item How can the countermeasure network actually help building a secure speaker verification against spoofing?
\end{itemize}
To answer these questions, we first study feature spaces that, when spoofed speeches are projected, reveals their artifacts more clearly.
After that, we examine whether there is no performance degradation by replacing the back-end model with a simple one upon using the feature extractor.
We further propose a simple system inspired by representation distillation \cite{tian2019contrastive} and self-distillation \cite{zhang2019your} for spoof-aware speaker verification utilizing the improved countermeasure system.


\section{Methods}
\label{sec:methods}

\subsection{Self-supervised representation for spoofing detection}
\label{ssec:ssl-cm}
To explore an effective representation space for spoofing detection, we divide the CM system into two parts, front-end and back-end.
We define the front-end by the part that projects the audio input to the feature space, and the back-end by the rest of the network that generates the embeddings for the final classification.
We study the feature space for spoofing countermeasure, using a pre-trained wav2vec 2.0 model as a front-end.
The architecture of wav2vec 2.0 consists of a convolutional encoder to map raw audio to latent speech representations and stacks of Transformer blocks that build context representations \cite{baevski2020wav2vec}.

The wav2vec model is trained to predict speech representations by solving a contrastive task, which requires identifying the quantized latent speech representation of the convolutional encoder for a time step within a set of distractors.
Through this self-supervised framework, the model builds context representations by capturing dependencies over the entire sequence of latent representations without any labels.
Such context can be correlated with various features including audial features, fluency, pronunciation, or semantic level text features \cite{shah2021all}.
Also, it has empirically been observed that, in controlled settings of clean and native speech inputs, wav2vec 2.0 model outperforms Mockingjay \cite{liu2020mockingjay} and BERT \cite{devlin2018bert} on audio and fluency feature representations \cite{shah2021all}.
Assuming the attacks are attempted using clean native speeches, we fix wav2vec 2.0 pre-trained model and examine whether each Transformer block's output serves as an effective feature representation for the countermeasure network.

In addition, we experimented with how lightweight the back-end model can be underneath the feature space of each front-end.
Through this experiment, we analyze how advantageous each feature extractor is in distinguishing artifacts, even though the constraints for the back-end architecture are considerably relaxed.
We adopt AASIST as a baseline of the back-end model, which is equipped with a front-end based on SincNet \cite{ravanelli2018speaker} and a back-end based on heterogeneous stacking graph attention mechanism \cite{jung2021aasist}.
Unless otherwise stated, we denote AASIST to mean only the back-end system of \cite{jung2021aasist}.
As simple back-end models, we examine a 3-layer multilayer perceptron (MLP), and an attentive statistics pooling (ASP) layer \cite{desplanques2020ecapa} equipped with a fully-connected (FC) layer.
The attentive statistics pooling layer calculates the mean and standard deviations of the front-end features and allows the model to select the frames that are deemed to be relevant via the attention mechanism.

\subsection{Spoof-aware speaker verification}
Our SASV system consists of three neural networks: a speaker verification network $E$ based on ECAPA-TDNN \cite{desplanques2020ecapa}, a spoofing countermeasure network $C$ based on section \ref{ssec:ssl-cm}, and a newly proposed representation selective self-distillation (RSSD) module based on representation distillation.
While we keep $E$ and $C$ to be independently trained, RSSD is trained to selectively transform the speaker representations by $E$ using the countermeasure embeddings of $C$.

Let $\mathbf{e}_e=E(\mathbf{x}_e)$ and $\mathbf{e}_t=E(\mathbf{x}_t)$ denote speaker embeddings of enrollment and test speaker, respectively.
Denote countermeasure embedding of the test utterance by $\mathbf{c}_t=C(\mathbf{x}_t)$.
We measure spoof-aware speaker similarity by
\begin{equation}\label{eqn:sim}
    \mathrm{sim}(\mathbf{e}_e, g(f_\theta(\mathbf{c}_t), \mathbf{e}_t)),
\end{equation}
where $\mathrm{sim}(\mathbf{a},\mathbf{b})=\langle \mathbf{a},\mathbf{b}\rangle/\|\mathbf{a}\|\|\mathbf{b}\|$ denotes cosine similarity.
RSSD network consists of two modules: a feature transform layer $f_\theta$ and a gating operator $g$.
The layer $f_\theta$ transform the embedding $\mathbf{c}_t$ of test speech $\mathbf{x}_t$, to modulate the speaker embedding of $\mathbf{x}_t$.
Then, gate operator $g$ outputs self-distilled representation using the two embeddings $f_\theta(\mathbf{c}_t)$ and $\mathbf{e}_t$.
Finally, we measure spoof-aware speaker similarity as equation \eqref{eqn:sim}.

During training, our objective function $\mathcal{L}$ is computed for given triad of labeled utterances $(\mathbf{x}_e, \mathbf{x}_t,y)$ as
\begin{equation}\label{eqn:total}
    \mathcal{L}(\mathbf{x}_e, \mathbf{x}_t) = \mathbbm{1}_{[y=1]} \mathcal{L}_{\text{r-distill}}(\mathbf{x}_t) + \mathbbm{1}_{[y\neq1]} \mathcal{L}_{\text{spoof}}(\mathbf{x}_e, \mathbf{x}_t),
\end{equation}
where the label $y=1$ if $\mathbf{x}_t$ is bona fide, and $y=0$ otherwise.
$\mathcal{L}_{\text{r-distill}}$ denotes representation self-distillation loss, and $\mathcal{L}_{\text{spoof}}$ denotes spoofing countermeasure loss.

\subsubsection{Representation self-distillation loss}
RSSD distills the information through representations extracted from the pre-trained speaker verification network.
This is the difference from conventional knowledge distillation which distill the information through probabilistic outputs from teacher networks \cite{hinton2015distilling}.
The representation self-distillation loss $\mathcal{L}_{\text{r-distill}}$ of bona fide example $\mathbf{x}$ is defined as
\begin{equation}
    \mathcal{L}_{\text{r-distill}}(\mathbf{x}_t) = - \mathrm{sim}(\mathbf{e}_t, g(f_\theta(\mathbf{c}_t),\mathbf{e}_t)),
\end{equation}
where $\mathbf{e}_t = E(\mathbf{x}_t)$ denotes representation from the pre-trained speaker verification network, and $\mathbf{c}_t = C(\mathbf{x}_t)$ denotes representation from the pre-trained countermeasure network.
Parameters of the feature transformation layer $f_\theta(\cdot)$ are optimized to preserve the bona fide speaker representation from the pre-trained network.
The transformed countermeasure representation $f_\theta(C(\cdot))$ determines whether to preserve speaker representation $E$ using the gating operator $g$ or not.

\subsubsection{Spoofing countermeasure loss}
The spoofing loss $\mathcal{L}_{\text{spoof}}$ between spoof example $\mathbf{x}_t$ and enrollment example $\mathbf{x}_e$ is defined as
\begin{equation}
    \mathcal{L}_{\text{spoof}}(\mathbf{x}_e,\mathbf{x}_t) = \mathrm{sim}(\mathbf{e}_e, g(f_\theta(
    \mathbf{c}_t), \mathbf{e}_t)),
\end{equation}
where $\mathbf{e}_e = E(\mathbf{x}_e)$ denotes speaker representation of the enrollment speaker.
In contrast to bona fide examples, the desired representation space should not distill the information of given spoof speaker representations so that transformed representation $g(f_\theta(\mathbf{c}_t), \mathbf{e}_t))$ drifts apart from original enrollment speaker representation $\mathbf{e}_e$.
The feature transformation layer $f_\theta$ changes spoof countermeasure representation $\mathbf{c}_t$ to displace spoof speaker representation far from enrollment speaker representation.

\begin{figure*}[t]
    \centering
    \setlength{\abovecaptionskip}{0pt}
    \setlength{\belowcaptionskip}{-15pt}
    \caption{Breakdown of CM-EERs tested using XLSR-53 front-end with ASP back-end, for the spoofing countermeasure task.
    }
    \centerline{\includegraphics[width=\linewidth]{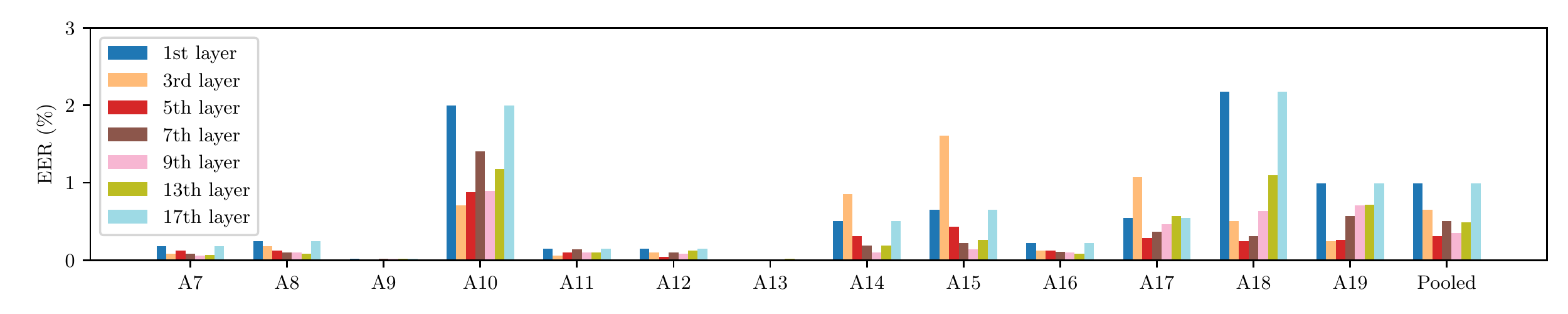}}
    \label{fig:breakdown-layerwise-asp}
\end{figure*}

\section{Experimental Setup}
\label{sec:experiment}


\subsection{Implementation details}
\label{ssec:details}

For all of the experiments, we use ASVspoof 2019 LA database \cite{todisco2019asvspoof}.
It consists of bona fide speeches collected from VCTK corpus \cite{yamagishi2019cstr}, and spoofed speeches generated using a variety of TTS and VC methods.
The training and development sets were constructed using 6 different algorithms, while the evaluation set consists of non-overlapping 13 different methods \cite{wang2020asvspoof}.
The training, development, and evaluation datasets consist of 20, 20, and 67 unique speakers, respectively.
For more specific details on the experimental dataset, we refer to \cite{todisco2019asvspoof,wang2020asvspoof}.

\subsubsection{Feature exploration of spoofing countermeasure}
\label{ssec:spf}
We conduct experiments to replace the front-end of AASIST by XLSR-53 \cite{conneau2020unsupervised}: a wav2vec 2.0 model pre-trained on 56k hours of speech in 53 languages.
We use the XLSR-53 Large model implemented in fairseq\footnote{\url{https://github.com/pytorch/fairseq/tree/main/examples/wav2vec}} \cite{ott2019fairseq}, equipped with 24 Transformer blocks with dimension of 1024.
For the back-end model comparison, we use an MLP that consists of 3 FC layers with dimension 1024 and leaky ReLUs between each layer.
We also examine the use of ASP layer that pools the feature representation into statistics of dimension 2048.
The statistics are linearly projected to represent an embedding of dimension 160.

\newcommand{\cmark}{\ding{51}}%
\newcommand{\xmark}{\ding{55}}%

\begin{table}[t]
    \centering
    \caption{CM-EERs for spoofing countermeasure networks.
    All self-supervised (SS) front-ends below consist of 317M number of parameters.
    `Inter.' indicates whether an intermediate feature is used, other than the final output of Transformer.
    }
    \footnotesize
    \begin{tabular}{ll ccc}
        \toprule
        \textbf{System} & \textbf{Front-end} &
        \textbf{SS} & \textbf{Inter.} & \textbf{EER}
        \\
        \midrule
        RawGAT-ST \cite{tak2021endGAT} &
        SincNet \cite{ravanelli2018speaker} &
        \xmark & - & 1.06
        \\
        AASIST-L \cite{jung2021aasist} &
        SincNet  \cite{ravanelli2018speaker} &
        \xmark & - & 0.99
        \\
        AASIST \cite{jung2021aasist} &
        SincNet \cite{ravanelli2018speaker} &
        \xmark & - & 0.83
        \\
        LGF \cite{wang2021investigating} &
        XLSR-53 \cite{conneau2020unsupervised} &
        \cmark & \xmark & 1.28
        \\
        LLGF \cite{wang2021investigating} &
        W2V-Large \cite{baevski2020wav2vec} &
        \cmark & \xmark & 0.86
        \\
        \midrule
        Ours (MLP) &
        XLSR-53 \cite{conneau2020unsupervised} &
        \cmark & \cmark & 0.80
        \\
        Ours (AASIST) &
        XLSR-53 \cite{conneau2020unsupervised} &
        \cmark & \cmark & 0.40
        \\
        Ours (ASP) &
        XLSR-53 \cite{conneau2020unsupervised} &
        \cmark & \cmark & \textbf{0.31}
        \\
        \bottomrule
    \end{tabular}
    \label{tab:fe-comparison}
\end{table}

\subsubsection{Spoof-aware speaker verification}
We use 3 fully connected layers with leaky ReLU activation for $f_\theta$ and Hadamard product for the gating operator $g$.
We train RSSD with batch size 32 using Adam optimizer \cite{kingma2014adam} for 20 epochs with the learning rate $10^{-4}$.
For evaluation, we use a model with the best scores on the development set.
We adopt two methods presented in the SASV challenge \cite{jung2022sasv} as our baselines.
As with the baselines, we also leverage the pre-trained model of state-of-the-art ASV network ECAPA-TDNN\footnote{\url{https://huggingface.co/speechbrain/spkrec-ecapa-voxceleb}}.

\subsection{Evaluation metric}
\label{ssec:metric}
We use EER (\%) as the evaluation metric for every experiment, where we denote the EER of the CM and ASV tasks as CM-EER and SV-EER, respectively.
On the SASV task, we identify the EER for the spoofed input as SPF-EER.
The performance of RSSD is evaluated using SASV-EER \cite{jung2022sasv}, which is a combination of SPF-EER and SV-EER.
SASV-EER shows how well a system can distinguish bona fide and target signals from spoofed or (zero-effort) non-target ones.

\section{Results}
\label{sec:result}

\subsection{Spoofing countermeasure}
Table \ref{tab:fe-comparison} compares EERs of CM systems with different combinations of front-end and back-end modules.
For the same back-end system with AASIST, we achieve a 43\% performance improvement just by replacing the front-end from SincNet to XLSR-53.
Moreover, a simple 3-layer MLP back-end with XLSR-53 front-end performed even better than the original AASIST.
Yet, this observation involves a more in-depth analysis than simply replacing the front-end with a self-supervised model.
Compared to the best-reported performance of \cite{wang2021investigating}, where they used the output of the final Transformer layer of XLSR-53 or wav2vec 2.0 Large model, our approach of selecting an intermediate, 5\textsuperscript{th}, layer unveils the effective feature space for CM by outperforming the previous studies.
This result suggests the importance of exploring the feature space of each wav2vec 2.0 layer.

\subsubsection{Feature exploration}
Figure \ref{fig:breakdown-layerwise-asp} shows EERs of our CM system with XLSR-53 front-end and ASP back-end.
Among 24 Transformer layers of XLSR-53, using features from 5\textsuperscript{th} layer shows the lowest pooled EER.
This observation can be interpreted by connecting with the empirical analysis of \cite{shah2021all} that audial features (e.g., local jittering or shimmering) are highly related to the features from 2\textsuperscript{nd} out of 12 Transformer layers.
Hence, it can be inferred that synthetic artifacts are more highly exposed by the audial features than other characteristics such as fluency, pronunciation, and text semantics, where \cite{shah2021all} reported they are best learned from the 6\textsuperscript{th} to 9\textsuperscript{th} layers.

Our system shows strength against TTS attacks (A07-12, A16) compared to VC attacks (A13-15, A17-19), except for A10.
This can be attributed to vocoders that also synthesized spoofed training data or are difficult to produce utterances sufficiently similar to genuine speech in the current setting \cite{wang2020asvspoof}.
A10 was relatively more difficult to figure out regardless of back-end models, which is similar to the report of \cite{wang2020asvspoof} that the difference between A10 and bona fide utterances is marginally significant.
A10 is the only approach to use WaveRNN \cite{kalchbrenner2018efficient} vocoder.

For VC attacks, we observe that EER can vary depending on the quality of the input source rather than the vocoder.
We could easily identify A13 which used a publicly available TTS system as the input source of the VC system, but relatively higher EER was scored for A14 and A15 which are of similar approach but used a commercial TTS synthesis engine.
The vocoder of A15 is the same as that used for the training data A01 which was included in the training data for the back-end.
A13 and A17 use the same vocoder which is based on waveform filtering, but the EER of A17, which receives genuine human voice as a conversion source, is measured higher.

Table \ref{tab:layerwise} compares CM-EERs for various front-ends.
For all choices of back-end models, using 5\textsuperscript{th} layer of XLSR-53 achieved the best performance.
We observe that a single ASP layer outperforms AASIST, in the case of using XLSR-53 as the front-end.
It can be inferred that the attentive pooling of self-supervised representations is effective for spoofing detection.

\begin{table}[t]
    \centering
    \setlength{\abovecaptionskip}{5pt}
    \caption{Pooled CM-EERs for CM task.
    The result of the model with sinc-AASIST is taken from the best EER reported in \cite{jung2021aasist}.
    All EERs are rounded to the first decimal place.
    }
    \footnotesize
    \begin{tabular}{l c ccccccc}
        \toprule
        \multirow{2}{*}{\textbf{Back-end}}
        & \multirow{2}{*}{sinc} & \multicolumn{7}{c}{XLSR-53} \\\cmidrule{3-9}
        & & 
        1\textsuperscript{st} & 3\textsuperscript{rd} &
        5\textsuperscript{th} & 7\textsuperscript{th} &
        9\textsuperscript{th} & 13\textsuperscript{th} &
        17\textsuperscript{th} \\
        \midrule
        MLP     &
        - & 
        2.2 & 2.1 & 0.8 & 1.4 & 1.0 & 1.7 & 2.2
        \\
        AASIST  &
        0.8 &
        2.9 & 0.8 & 0.4 & 0.5 & 0.5 & 0.7 & 0.9
        \\
        ASP &
        1.0 &
        1.0 & 0.7 & 0.3 & 0.5 & 0.4 & 0.5 & 1.0
        \\
        \bottomrule
    \end{tabular}
    \label{tab:layerwise}
\end{table}

\begin{figure}[b]
    \centering
    \setlength{\abovecaptionskip}{0pt}
    \centerline{\includegraphics[width=\linewidth]{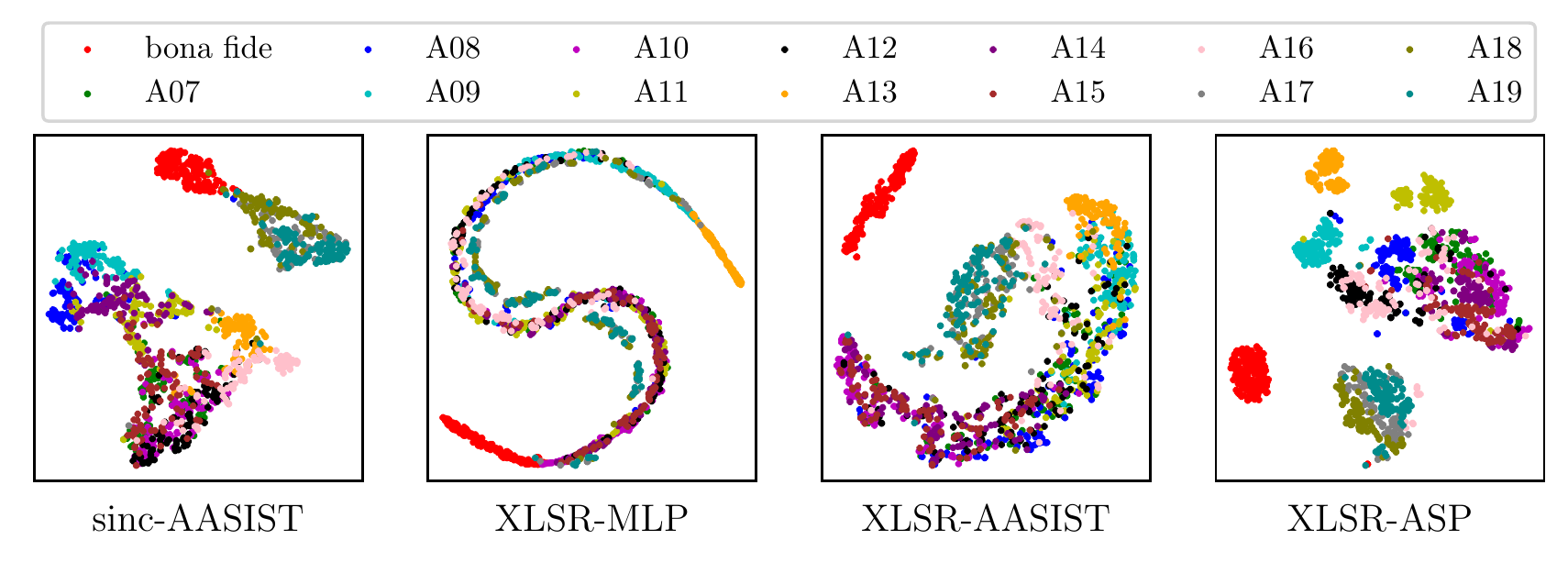}}
    \caption{t-SNE samples of countermeasure embeddings for different types of spoofing attacks in the evaluation set.
    The caption below each image indicates the front and back-end pair used to extract the embeddings.
    The scattered embeddings are displayed in different colors depending on the attack type, and the embeddings for the bona fide utterances are marked in red.
    }
    \label{fig:tsne}
\end{figure}

\subsubsection{Back-end model selection for self-supervised front-ends}
Scatter plots of t-SNE for CM embeddings are illustrated in Figure \ref{fig:tsne}.
For the original AASIST with sinc front-end, we observe that the embeddings corresponding to A17 (gray) and A18 (olive) are entangled with those of bona fide (red) utterances.
Considering that the two attack types which recorded the highest EER of AASIST are A17 and A18 \cite{jung2021aasist}, this observation can be considered to be consistent with the breakdown EER of original AASIST.
Embeddings extracted using XLSR-53 as front-end show better disentanglement of bona fide utterances against A17 and A18.

Above all, ASP with XLSR-53 front-end shows the best disentanglement.
Although no information on attack types was included in the training, ASP even distinguishes the embeddings for attack types A09, A11, and A13.
A13 is known to deceive CM systems relatively well, even though it is easily identified by humans \cite{wang2020asvspoof}, and we can easily distinguish it down to 0.00\% EER.
Interestingly, the embeddings corresponding to the VC systems (A17-19) are clustered together (bottom of the XLSR-ASP t-SNE).
Similar to the observation of \cite{wang2020asvspoof}, VC systems that converted synthetic speeches (A13-15) tend to be grouped with TTS systems.

Table \ref{tab:layerwise} shows that with help of an appropriate choice of feature extractor, the performance could outperform the existing system even with a simpler back-end system.
For all backend systems, using the output of the 5\textsuperscript{th} layer as a feature show the best performance.
Also, we find that changing the back-end of AASIST to a 3-layer MLP did not cause significant performance degradation under the appropriate front-end selection.
However, even with a fewer number of parameters than the rest, ASP effectively improves the countermeasure, by even outperforming the others for most front-ends.
It can be inferred that with a suitable choice of feature extractor, attentive pooling operation can outperform other back-ends.

\subsection{Spoof-aware speaker verification}
Table \ref{tab:model-comparison} compares our method with the SASV 2022 challenge baselines.
By replacing the baseline SASV systems with RSSD using the original AASIST as CM network, we achieve 1.15\% SASV-EER, surpassing the baselines by a large margin.
We further increase the performance of RSSD by replacing the front-end of the CM network with XLSR-53, and achieve the best performance of 1.08\% SASV-EER by adopting ASP as a back-end.
It is also worth mentioning that ECAPA-TDNN achieves 1.63\% SV-EER for ASV task and the original AASIST achieves 0.83\% CM-EER for CM task.
Since RSSD outperforms each system by solving SASV task, it can be inferred that RSSD not only distills the representations of the subsystems, but also completes the embedding space to improve the performance of the subtasks.

\begin{table}[t]
    \centering
    \setlength{\abovecaptionskip}{5pt}
    \footnotesize
    \caption{Pooled EERs with RSSD for SASV task.
    For the same ASV and CM subsystem, RSSD significantly outperforms the baselines.
    Improving the CM network further develops RSSD.
    }
    \begin{tabular}{l ll rrr}
        \toprule
        \textbf{System} &
        \textbf{CM Front} & \textbf{CM Back} &
        \textbf{SV} & \textbf{SPF} & \textbf{SASV} \\
        \midrule
        Baseline 1 & 
        SincNet & AASIST &
        35.32 & 0.67 & 19.31
        \\
        Baseline 2 & 
        SincNet & AASIST &
        11.48 & 0.78 & 6.37
        \\
        \midrule
        RSSD &
        SincNet & AASIST &
        {1.41}  & 0.76 & {1.15}
        \\
        RSSD &
        XLSR-53 & AASIST &
        {1.34} & 0.60 & {1.11}
        \\
        RSSD &
        XLSR-53 & ASP &
        {1.32} & 0.59 & {1.08}
        \\
        \bottomrule
    \end{tabular}
    \label{tab:model-comparison}
\end{table}

\section{Conclusion}
\label{sec:conclusion}
This paper explores the use of wav2vec 2.0 as the feature extractor of the spoofing countermeasures.
We investigate from which layer of XLSR-53 the features obtained are advantageous for defending spoof attacks.
Quantitative analysis shows that using the features of the 5\textsuperscript{th} layer of XLSR-53 is effective for spoofing detection, outperforming the existing sinc convolution-based front-end model.
We further find that for such feature representations, the back-end system with a simple ASP layer outperformed AASIST and MLP.
With the findings of wav2vec 2.0 features for spoofing detection, we propose RSSD
for SASV to fully utilize the state-of-the-art models from both ASV and spoofing detection.
By selectively performing speaker verification based on self-distillation using spoof countermeasures and speaker embeddings, RSSD achieves SASV-EER of 1.08\%.
Empirical results show that RSSD takes advantage of the state-of-the-art speaker verification and spoofing detection models with efficient time and memory complexity.

\section{Acknowledgement}

This work was partly supported by
Institute of Information \& communications Technology Planning \& Evaluation (IITP) grant funded by the Korea government(MSIT) (No. 2022-0-00320, Artificial intelligence research about cross-modal dialogue modeling for one-on-one multi-modal interactions, 90\%)
and
[NO.2021-0-01343, Artificial Intelligence Graduate School Program (Seoul National University), 10\%].

\newpage
\bibliographystyle{IEEEtran}
\bibliography{template}

\newpage
\appendix
\onecolumn

\section{Countermeasure analysis results}
\begin{figure}[h]
    \begin{minipage}[b]{0.33\linewidth}
        \centering
        \centerline{\includegraphics[width=\linewidth]{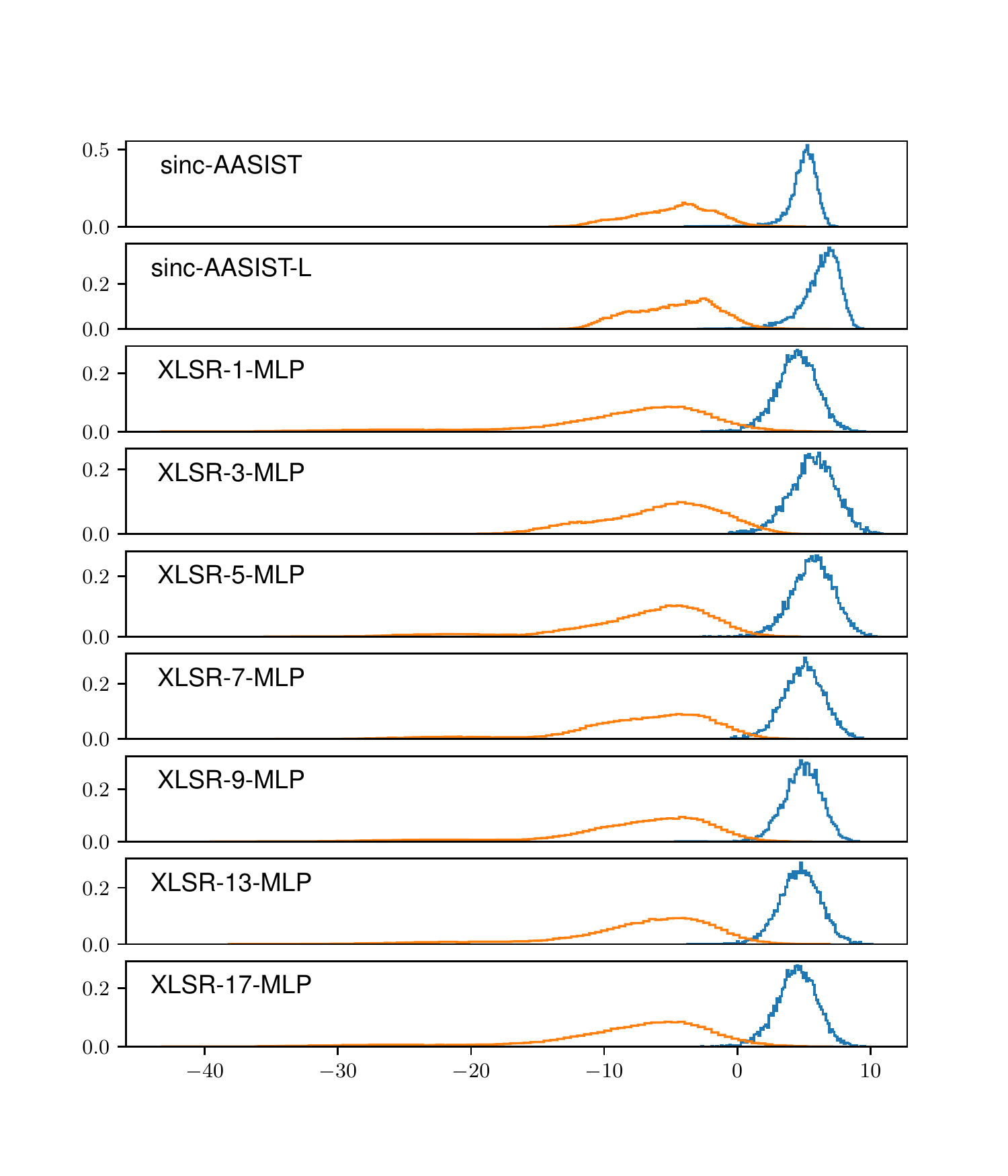}}
        \centerline{(a) XLSR-MLP}\medskip
    \end{minipage}
    \begin{minipage}[b]{0.33\linewidth}
        \centering
        \centerline{\includegraphics[width=\linewidth]{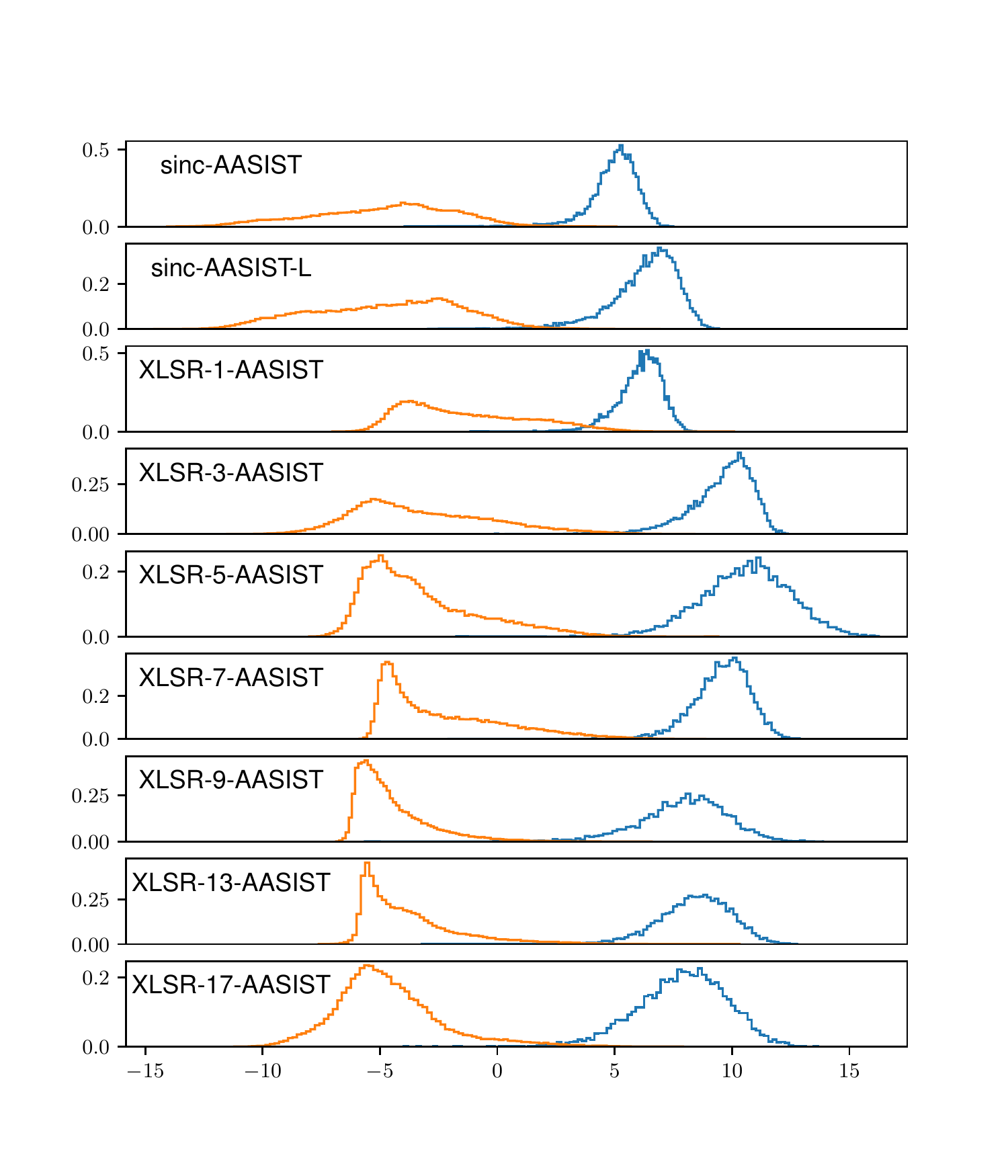}}
        \centerline{(b) XLSR-AASIST}\medskip
    \end{minipage}
    \begin{minipage}[b]{0.33\linewidth}
        \centering
        \centerline{\includegraphics[width=\linewidth]{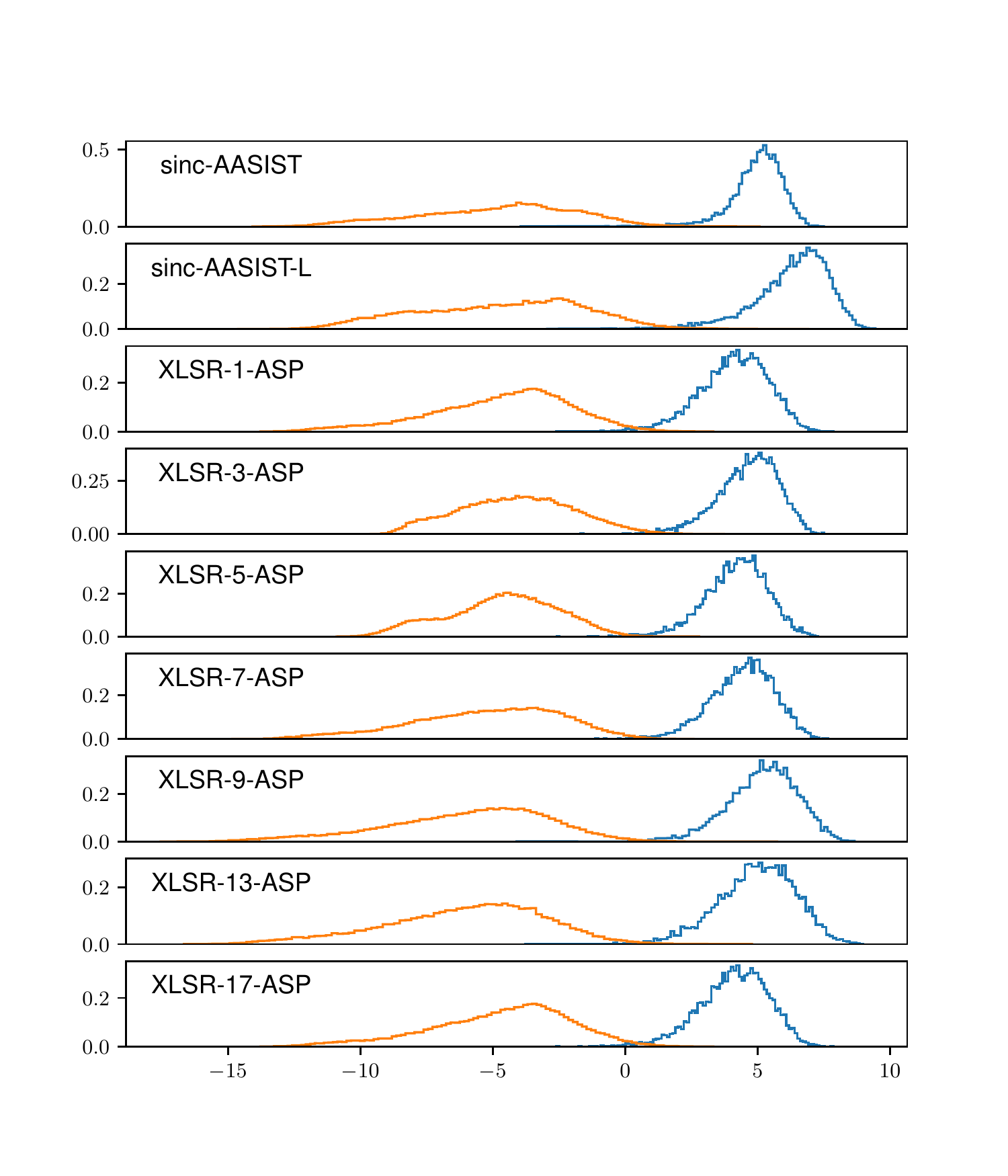}}
        \centerline{(c) XLSR-ASP}\medskip
    \end{minipage}
    \begin{minipage}[b]{\linewidth}
        \centering
        \centerline{\includegraphics[width=.5\linewidth]{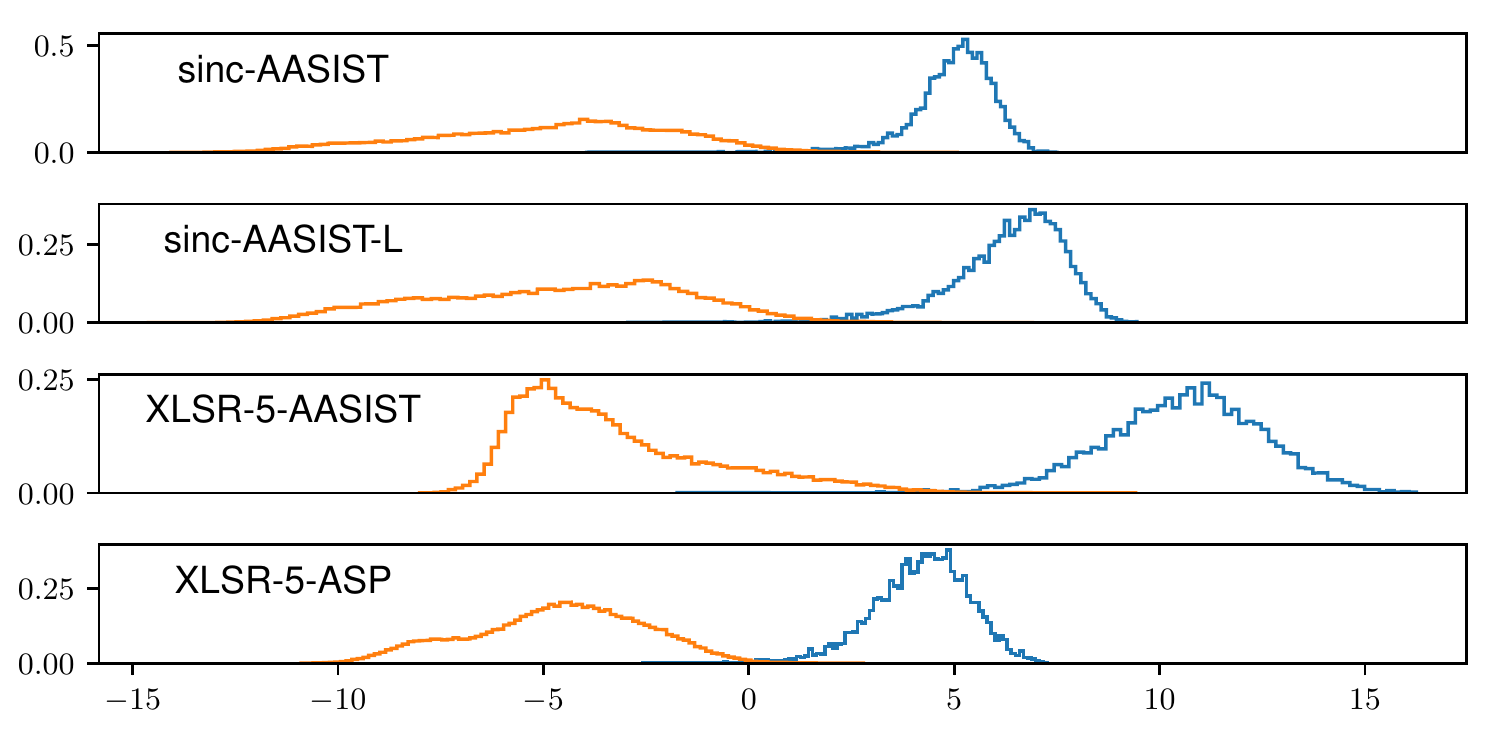}}
        \centerline{(d) Using 5th Transformer layer output}\medskip
    \end{minipage}
    \caption{Score distribution of various back-end models for each Transformer layer of XLSR-53 front-end.}
    \label{fig:score-full}
\end{figure}

\begin{figure*}[h]
    \centering
    \begin{minipage}[b]{.9\linewidth}
        \centerline{\includegraphics[width=\linewidth]{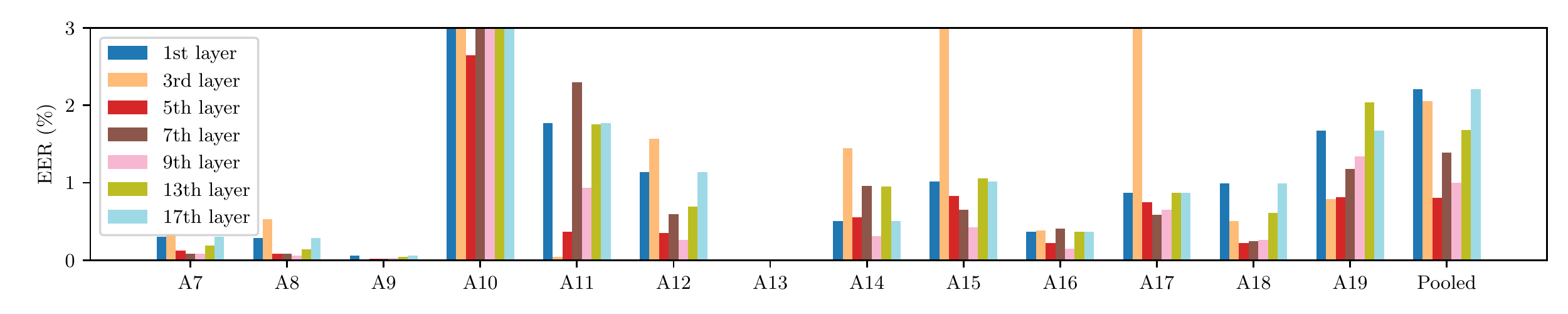}}
        \centerline{(a) XLSR-MLP}\medskip
    \end{minipage}
    \begin{minipage}[b]{.9\linewidth}
        \centerline{\includegraphics[width=\linewidth]{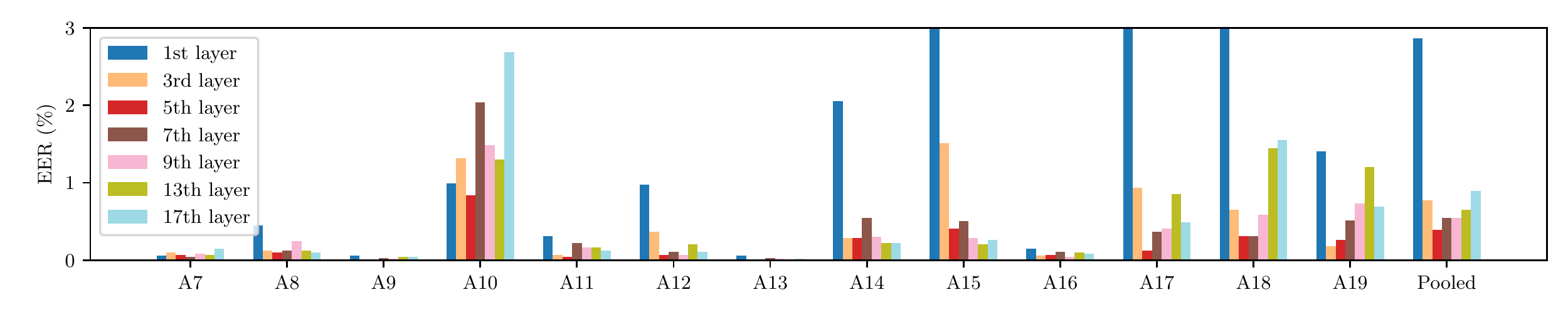}}
        \centerline{(b) XLSR-AASIST}\medskip
    \end{minipage}
    \begin{minipage}[b]{.9\linewidth}
        \centerline{\includegraphics[width=\linewidth]{fig/breakdown-layerwise-asp.pdf}}
        \centerline{(c) XLSR-ASP}\medskip
    \end{minipage}
    \caption{Breakdown of CM-EERs tested using XLSR-53 front-end with various back-ends. Among 24 layers of XLSR-53, using features from front level layers such as 5th, or 9th showed low EERs against most attacks.
    For attack type A10, relatively high EERs were measured for most layers.
    For overall visibility, values higher than 3\% were clamped while plotting.
    We provide the breakdown EER scores in Table \ref{tab:bd-mlp}, \ref{tab:bd-aasist}, and \ref{tab:bd-asp}.
    }
    \label{fig:breakdown-layerwise-full}
\end{figure*}

We show more diverse experimental results with respect to countermeasure feature space exploration and the corresponding back-end model.
\begin{itemize}
    \item Figure \ref{fig:score-full} compares the score distribution for various back-end models, throughout the XLSR-53 Transformer layers.
    \item Figure \ref{fig:score-full} (d) comprares the score distribution for some models appearing in Table \ref{tab:fe-comparison}.
    \begin{itemize}
        \item The distribution of scores between AASIST and AASIST-L using the SincNet front-end shows little difference.
        \item Replacing SincNet with XLSR-53 makes the score for the spoofed sample appear more skewed.
        \item Selecting ASP as the back-end model can affect shortening the tail of the non-target sample distribution, resulting in the reduced EER.
    \end{itemize}
    \item Figure \ref{fig:breakdown-layerwise-full} shows the breakdown EERs for various back-end models, with XLSR-53 front-end.
    The performance that varies according to the layer as well as the architecture of the back-end model motivates the feature space exploration.
\end{itemize}

\begin{itemize}
    \item Table \ref{tab:breakdown} shows the breakdown EERs using various front-end and back-end pairs of CM system.
    \begin{itemize}
        \item Among the results shown in Table \ref{tab:breakdown}, all those calculated using XLSR-53 as front-end used the 5th layer output.
        \item We also compare the results of the original AASIST (denoted by sinc-AASIST).
        \item Using XLSR-53 front-end significantly advantages detecting the spoofing attack of type A13, regardless of the choice of the back-end model.
        \item For A10, which is the attack that our system is most vulnerable to, performance similar to (or superior to) the original AASIST appears depending on the selection of the back-end model.
    \end{itemize}
    \item Table \ref{tab:model-comparison-dev} shows the same result as Table \ref{tab:model-comparison}, but with the results for the development set.
    \item Table \ref{tab:bd-mlp}, \ref{tab:bd-aasist}, and \ref{tab:bd-asp} show the breakdown EERs for the front-end using more diverse Tranformer layer outputs, with the back-end of MLP, AASIST, and ASP, respectively.
\end{itemize}

\begin{table*}[t]
    \centering
    \caption{Breakdown SPF-EER (\%) performance for systems using features extracted from 5\textsuperscript{th} layer of XLSR-53.
    }
    \footnotesize
    \begin{tabular}{l ccccc ccccc ccc cc}
        \toprule
        \textbf{CM System} &
        \textbf{A07} &\textbf{A08} &\textbf{A09} &\textbf{A10} &\textbf{A11} &
        \textbf{A12} &\textbf{A13} &\textbf{A14} &\textbf{A15} &\textbf{A16} &
        \textbf{A17} &\textbf{A18} &\textbf{A19} &
        \textbf{Pooled} & \textbf{min t-DCF} \\
        \midrule
        sinc-AASIST &
        0.53 & 0.42 & \textbf{0.00} & 0.86 & 0.18 & 0.71 & 0.15 & \textbf{0.16} & 0.55 & 0.65 & 1.26 & 2.61 & 0.65 & 0.83 &
        0.0275
        \\
        XLSR-MLP     &
        0.12 &	\textbf{0.08} &	0.02 &	2.65 &	0.37 &	0.35 &	\textbf{0.00} &	0.55 &	0.83 &	0.22 &	0.75 &	\textbf{0.22} &	0.81 &	0.80 &
        0.0222
        \\
        XLSR-AASIST  &
        \textbf{0.06} &	0.10 &	\textbf{0.00} &	\textbf{0.84} &	\textbf{0.04} &	{0.06} &	\textbf{0.00} &	{0.29} &	\textbf{0.41} &	\textbf{0.06} &	\textbf{0.12} &	0.31 &	\textbf{0.26} &	{0.40} &
        {0.0100}
        \\
        XLSR-ASP &
        0.12 & 0.12 & \textbf{0.00} & 0.89 & 0.10 & \textbf{0.04} & \textbf{0.00} & 0.31 & 0.43 & 0.12 & 0.29 & 0.24 & \textbf{0.26} &
        \textbf{0.31} &
        \textbf{0.0088}
        \\
        \bottomrule
    \end{tabular}
    \label{tab:breakdown}
\end{table*}
\begin{table}[h]
    \centering
    \footnotesize
    \caption{Pooled EERs with RSSD for SASV task.
    This table shows more details about Table \ref{tab:model-comparison}.
    {\tt dev} and {\tt eval} denotes the development and evaluation sets of the database, respectively.
    }
    \begin{tabular}{l lll rr rr rr}
        \toprule
        \multirow{2}{*}{\textbf{SASV System}} &
        \multirow{2}{*}{\textbf{ASV System}} &
        \multicolumn{2}{c}{\textbf{CM System}} &
        \multicolumn{2}{c}{\textbf{SV-EER}} &
        \multicolumn{2}{c}{\textbf{SPF-EER}} &
        \multicolumn{2}{c}{\textbf{SASV-EER}}
        \\
        & & Front-end & Back-end &
        {\tt dev} & {\tt eval} &
        {\tt dev} & {\tt eval} &
        {\tt dev} & {\tt eval} \\
        \midrule
        Baseline 1 & ECAPA-TDNN \cite{desplanques2020ecapa} &
        SincNet \cite{ravanelli2018speaker} & AASIST \cite{jung2021aasist} &
        32.88 & 35.32 & 0.06 & 0.67 & 13.07 & 19.31
        \\
        Baseline 2 & ECAPA-TDNN \cite{desplanques2020ecapa} & 
        SincNet \cite{ravanelli2018speaker} & AASIST \cite{jung2021aasist} &
        12.87 & 11.48 & 0.13 & 0.78 & 4.85 & 6.37
        \\
        \midrule
        RSSD & ECAPA-TDNN \cite{desplanques2020ecapa} &
        SincNet \cite{ravanelli2018speaker} & AASIST \cite{jung2021aasist} &
        {2.01} & {1.41} & 0.17 & 0.76 & {1.16} & {1.15}
        \\
        RSSD & ECAPA-TDNN \cite{desplanques2020ecapa} &
        XLSR-53 \cite{conneau2020unsupervised} & AASIST \cite{jung2021aasist} &
        {1.88} & {1.34} & 0.13 & 0.60 & {1.07} & {1.11}
        \\
        RSSD & ECAPA-TDNN \cite{desplanques2020ecapa} &
        XLSR-53 \cite{conneau2020unsupervised} & ASP \cite{desplanques2020ecapa} &
        {1.95} & {1.32} & 0.13 & 0.59 & {1.07} & {1.08}
        \\
        \bottomrule
    \end{tabular}
    \label{tab:model-comparison-dev}
\end{table}

\begin{table}[h]
    \centering
    \caption{Breakdown EER (\%) performance of our model with MLP backend tested for all 13 attacks in the ASVspoof 2019 LA evaluation set.}
    \begin{tabular}{c ccccc ccccc ccc c}
        \toprule
        \textbf{Layer} &
        \textbf{A07} &\textbf{A08} &\textbf{A09} &\textbf{A10} &\textbf{A11} &
        \textbf{A12} &\textbf{A13} &\textbf{A14} &\textbf{A15} &\textbf{A16} &
        \textbf{A17} &\textbf{A18} &\textbf{A19} &
        \textbf{Pooled} \\
        \midrule
        1\textsuperscript{st} &
        0.30 &	0.29 &	0.06 &	10.48 &	1.77 &	1.14 &	0.00 &	0.51 &	1.02 &	0.37 &	0.87 &	0.99 &	1.67 &	2.21	\\
        3\textsuperscript{rd} &
        0.34 &	0.53 &	0.00 &	3.50 &	0.04 &	1.57 &	0.00 &	1.44 &	5.17 &	0.38 &	3.33 &	0.51 &	0.79 &	2.05	\\
        5\textsuperscript{th} &
        0.12 &	0.08 &	0.02 &	2.65 &	0.37 &	0.35 &	0.00 &	0.55 &	0.83 &	0.22 &	0.75 &	0.22 &	0.81 &	0.80	\\
        7\textsuperscript{th} &
        0.08 &	0.08 &	0.02 &	4.60 &	2.30 &	0.59 &	0.00 &	0.96 &	0.65 &	0.41 &	0.59 &	0.24 &	1.18 &	1.39	\\
        9\textsuperscript{th} &
        0.08 &	0.06 &	0.02 &	3.89 &	0.94 &	0.26 &	0.00 &	0.31 &	0.42 &	0.15 &	0.65 &	0.26 &	1.34 &	1.00	\\
        13\textsuperscript{th} &
        0.19 &	0.14 &	0.04 &	6.41 &	1.75 &	0.69 &	0.00 &	0.95 &	1.06 &	0.37 &	0.87 &	0.61 &	2.04 &	1.68	\\
        17\textsuperscript{th} &
        0.30 &	0.29 &	0.06 &	10.48 &	1.77 &	1.14 &	0.00 &	0.51 &	1.02 &	0.37 &	0.87 &	0.99 &	1.67 &	2.21	\\
        \bottomrule
    \end{tabular}
    \label{tab:bd-mlp}
\end{table}
\begin{table}[h]
    \centering
    \caption{Breakdown EER (\%) performance of our model with AASIST backend tested for all 13 attacks in the ASVspoof 2019 LA evaluation set.}
    \begin{tabular}{c ccccc ccccc ccc c}
        \toprule
        \textbf{Layer} &
        \textbf{A07} &\textbf{A08} &\textbf{A09} &\textbf{A10} &\textbf{A11} &
        \textbf{A12} &\textbf{A13} &\textbf{A14} &\textbf{A15} &\textbf{A16} &
        \textbf{A17} &\textbf{A18} &\textbf{A19} &
        \textbf{Pooled} \\
        \midrule
        1\textsuperscript{st} &
        0.06 &	0.45 &	0.06 &	0.99 &	0.31 &	0.98 &	0.06 &	2.05 &	4.31 &	0.15 &	10.64 &	3.13 &	1.40 &	2.87	\\
        3\textsuperscript{rd} &
        0.10 &	0.12 &	0.00 &	1.32 &	0.06 &	0.37 &	0.00 &	0.29 &	1.51 &	0.06 &	0.94 &	0.65 &	0.18 &	0.77	\\
        5\textsuperscript{th} &
        0.06 &	0.10 &	0.00 &	0.84 &	0.04 &	0.06 &	0.00 &	0.29 &	0.41 &	0.06 &	0.12 &	0.31 &	0.26 &	0.40	\\
        7\textsuperscript{th} &
        0.04 &	0.12 &	0.02 &	2.04 &	0.22 &	0.11 &	0.02 &	0.55 &	0.51 &	0.11 &	0.37 &	0.31 &	0.51 &	0.54	\\
        9\textsuperscript{th} &
        0.08 &	0.24 &	0.02 &	1.48 &	0.16 &	0.06 &	0.02 &	0.30 &	0.29 &	0.04 &	0.41 &	0.59 &	0.73 &	0.54	\\
        13\textsuperscript{th} &
        0.06 &	0.12 &	0.04 &	1.30 &	0.16 &	0.20 &	0.00 &	0.22 &	0.20 &	0.10 &	0.86 &	1.44 &	1.21 &	0.65	\\
        17\textsuperscript{th} &
        0.15 &	0.10 &	0.04 &	2.69 &	0.12 &	0.11 &	0.02 &	0.22 &	0.26 &	0.08 &	0.49 &	1.55 &	0.69 &	0.90	\\

        \bottomrule
    \end{tabular}
    \label{tab:bd-aasist}
\end{table}
\begin{table}[h]
    \centering
    \caption{Breakdown EER (\%) performance of our model with attentive statistics pooling backend tested for all 13 attacks in the ASVspoof 2019 LA evaluation set.}
    \begin{tabular}{c ccccc ccccc ccc c}
        \toprule
        \textbf{Layer} &
        \textbf{A07} &\textbf{A08} &\textbf{A09} &\textbf{A10} &\textbf{A11} &
        \textbf{A12} &\textbf{A13} &\textbf{A14} &\textbf{A15} &\textbf{A16} &
        \textbf{A17} &\textbf{A18} &\textbf{A19} &
        \textbf{Pooled} \\
        \midrule
        1\textsuperscript{st} &
        0.18 &	0.24 &	0.02 &	2.00 &	0.15 &	0.15 &	0.00 &	0.51 &	0.65 &	0.22 &	0.55 &	2.18 &	0.99 &	0.99	\\
        3\textsuperscript{rd} &
        0.08 &	0.18 &	0.00 &	0.71 &	0.06 &	0.10 &	0.00 &	0.86 &	1.61 &	0.12 &	1.08 &	0.51 &	0.24 &	0.65	\\
        5\textsuperscript{th} &
        0.12 &	0.12 &	0.00 &	0.88 &	0.10 &	0.04 &	0.00 &	0.31 &	0.43 &	0.12 &	0.29 &	0.24 &	0.26 &	0.31	\\
        7\textsuperscript{th} &
        0.08 &	0.10 &	0.02 &	1.40 &	0.14 &	0.10 &	0.00 &	0.19 &	0.22 &	0.11 &	0.37 &	0.31 &	0.57 &	0.50	\\
        9\textsuperscript{th} &
        0.06 &	0.10 &	0.02 &	0.90 &	0.10 &	0.08 &	0.00 &	0.10 &	0.14 &	0.10 &	0.47 &	0.63 &	0.71 &	0.35	\\
        13\textsuperscript{th} &
        0.06 &	0.08 &	0.02 &	1.18 &	0.10 &	0.12 &	0.02 &	0.19 &	0.26 &	0.08 &	0.57 &	1.10 &	0.72 &	0.49	\\
        17\textsuperscript{th} &
        0.18 &	0.24 &	0.02 &	2.00 &	0.15 &	0.15 &	0.00 &	0.51 &	0.65 &	0.22 &	0.55 &	2.18 &	0.99 &	0.99	\\
        \bottomrule
    \end{tabular}
    \label{tab:bd-asp}
\end{table}

\end{document}